\documentclass[11pt]{article}
\usepackage{amsmath,amssymb,amsfonts,fancyhdr,fancybox,graphics,epsfig,calc,color}
\renewcommand{\baselinestretch}{1.4}
\def\blist#1#2#3
{    \begin{list}{}{
     \setlength{\parsep}{0pt}
     \setlength{\leftmargin}{#1 pt}
     \setlength{\listparindent}{0pt}
     \setlength{\itemindent}{\listparindent}
     \setlength{\labelsep}{#3 pt}
     \setlength{\labelwidth}{\leftmargin}
     \addtolength{\labelwidth}{-\labelsep}
     \addtolength{\labelwidth}{\itemindent}
     \setlength{\rightmargin}{#2 pt}   }   }

\def\elist{\end{list}}

\newcommand{\bm}[1]{\mbox{\boldmath{$#1$}}}

\newcommand{\bbeta}{\boldsymbol\beta}
\newcommand{\bgamma}{\boldsymbol\gamma}
\newcommand{\btheta}{\boldsymbol\theta}
\newcommand{\bomega}{\boldsymbol\omega}
\newcommand{\bnu}{\boldsymbol\nu}

\newcommand{\by}{\bf y}

\renewcommand{\baselinestretch}{1.38}
\usepackage{amsmath}

\begin{document}

\topmargin -0.3in \oddsidemargin 0.2in \evensidemargin 0.2in
\baselineskip 9mm
\renewcommand \baselinestretch {1.2}
\title{A divergence formula for regularization methods with an $l_2$ constraint}
\date{}

\author{\begin{tabular}{c}
Yixin Fang$^1$\footnote{Correspondence to: yixin.fang@nyumc.org; 650 First Avenue, 5th floor,
New York, NY 10016, U.S.A.},\ \  Yuanjia Wang$^2$, and Xin Huang$^3$
\\\emph{$^1$Division of Biostatistics, School of Medicine, New York University}\\\emph{$^2$Department of Biostatistics, Columbia University}\\\emph{$^3$Public Health Sciences Division, Fred Hutchinson Cancer Research Center}
\end{tabular}}
\titlepage

\maketitle

\begin{center}
\begin{minipage}{130mm}
\begin{center}{\bf Abstract}\end{center}
We derive a divergence formula for a group of regularization methods with an $l_2$ constraint. The formula is useful for regularization parameter selection, because it provides an unbiased estimate for the number of degrees of freedom. We begin with deriving the formula for smoothing splines and then extend it to other settings such as penalized splines, ridge regression, and functional linear regression.

\end{minipage}
\end{center}


\newpage
\setcounter{equation}{0}

\section{Introduction}

A variety of regularization methods have been proposed in modern statistics (Hastie {\it et al.}, 2009). Usually, the regularization is controlled by a tuning parameter, and it is crucial to select an appropriate value of the tuning parameter. Many criteria have been discussed for tuning parameter selection (Hastie {\it et al.}, 2009). Some of these criteria, including AIC (Akaike, 1973),  GCV (Graven and Wahba, 1979), and BIC (Schwarz, 1978), depend on estimating the number of degrees of freedom, which measures the model complexity.

Here we consider the problem of estimating the number of degrees of freedom for a group of regularization methods, the ones with an $l_2$ constraints. We derive a divergence formula, which provides an unbiased estimate for the number of degrees of freedom (Stein, 1981; Ye, 1998; Efron, 2004). To understand our goal, we take smoothing slines as an example.

Smoothing splines are a popular approach to nonparametric function estimation (Wahba, 1990). Given observations
\begin{equation}
y_i=f(x_i)+\epsilon_i, \ \ i=1, \cdots, n, \label{model}
\end{equation}
where $x_i\in [0, 1]$ and $\epsilon_i\sim N(0, \sigma^2)$, one is to estimate $f(x)$. Assume that $f(x)$ is smooth in the sense that its second derivative exists and is small. The smoothing splines approach to the estimation of $f(x)$ is through minimizing
\begin{equation}
\sum_{i=1}^n(y_i-f(x_i))^2+\lambda\int_0^1\{f^{(2)}(x)\}^2dx, \label{penalty}
\end{equation}
where the {\it penalty parameter $\lambda$} controls the tradeoff between the lack of fit and the roughness of the function estimation. An alternative derivation is through minimizing
\begin{equation}
\sum_{i=1}^n(y_i-f(x_i))^2  {\mbox{\ \ s.t.\ }} \int_0^1\{f^{(2)}(x)\}^2dx\leq \rho, \label{constraint}
\end{equation}
where $\rho$ is called the {\it constraint parameter}. The solution to (\ref{constraint}) usually falls on the boundary of the constraint, and by the Lagrange method, these two formulations are equivalent up to the choice of $\lambda$ and $\rho$.

Gu (1998) pointed out that the mapping from $\rho$ to $\lambda$ is one-to-one but changes with the least squares functional, $\sum_{i=1}^n(y_i-f(x_i))^2$. In the literature, it is well-known that the divergence in terms of $\lambda$ is equal to the trace of the ``hat" matrix (e.g., Hastie and Tibshirani, 1990). However, the divergence formula in terms of $\rho$ is still not available. Because of the importance of tuning parameter selection, it is worth deriving this formula.

The remaining of the manuscript is organized as follows. In Section 2, the divergence formula in terms of $\rho$ is derived for smoothing splines. In Section 3, the result is extended to some other settings. In Section 4, the two divergence formulas for smoothing splines, in terms of $\lambda$ and $\rho$ respectively, are compared through a simulation study. Some discussion is in Section 5 and technical proof is in Appendix.

\section{Main Result}

\subsection{Definition}

We start with reviewing the definition of degrees of freedom. In ordinary linear regression, the degrees of freedom simply count the number of parameters. There are some generalizations of the degrees of freedom, including Stein's unbiased risk estimate in Stein (1981), the generalized degrees of freedom in Ye (1998), and the covariance penalty in Efron (2004).

Because the definitions are similar for both $\lambda$ and $\rho$, we use $\theta$ for the smoothing parameter, which could be either $\lambda$ or $\rho$. Let $\widehat{f}_{\theta}(\cdot)$ be the solution to either (\ref{penalty}) or (\ref{constraint}) for a given $\theta$, and then $\widehat{\mu}_i(\theta)=\widehat{f}_{\theta}(x_i)$ estimates $\mu_i=f(x_i)$. Assume that $y_i^0$ is a new response generated from the same mechanism that generates $y_i$. By arguments in Efron (2004), we obtain a decomposition of the prediction error,
\begin{eqnarray}
E\{\sum_{i=1}^n(y_i^0-\widehat{\mu}_i(\theta))^2\}=E\{\sum_{i=1}^n(y_i-\widehat{\mu}_i(\theta))^2\}+
2E\{{{\mbox{div}}}(\theta)\}\sigma^2,\label{prediction_error}
\end{eqnarray}
where
\begin{eqnarray}
{{\mbox{div}}}(\theta)=\sum_{i=1}^n\partial\widehat{\mu}_i(\theta)/\partial y_i \label{div}
\end{eqnarray}
is called the {\it divergence} in terms of $\theta$ (either $\lambda$ or $\rho$) in Kato (2009), and its expectation
\begin{eqnarray}
\mbox{DF}(\theta)=E\{{{\mbox{div}}}(\theta)\}\label{DF}
\end{eqnarray}
is defined as the {\it degrees of freedom} in terms of $\theta$ (either $\lambda$ or $\rho$). Clearly the divergence is an unbiased estimate of the degrees of freedom.

\subsection{The divergence in terms of $\lambda$}

Since the solution to (\ref{penalty}) is natural splines (Wahba, 1990), it can be written as
\begin{eqnarray}
f(x)=\sum_{j=1}^dN_j(x)\beta_j, \label{natural splines}
\end{eqnarray}
where $d=n$ in smoothing splines approach ($d$ could be different from $n$ in other settings), and $\{N_j(x)\}$ is an $n$-dimensional set of basis functions for representing this family of cubic natural splines. Then the regularization problem (\ref{penalty}) becomes
\begin{eqnarray}
\mbox{argmin\ }||\by-{\bf N}\bm\beta||^2_2+\lambda\bm\beta'\bm\Omega_N\bm\beta, \label{new_penalty}
\end{eqnarray}
where ${\by}=(y_1, \cdots, y_n)'$. $\bm\beta=(\beta_1, \cdots, \beta_d)'$, $\{{\bf N}\}_{ij}=N_j(x_i)$ and $\{\bm\Omega_N\}_{ij}=\int N_i^{(2)}(t)N_j^{(2)}(t)dt$.

The ``hat" matrix ${\bf H}(\lambda)={\bf N}({\bf N}'{\bf N}+\lambda\bm\Omega_N)^{-1}{\bf N}'\by$, and the divergence in terms of $\lambda$ is well-known to be equal to the trace of it (Hastie and Tibshirani, 1990), that is,
\begin{equation}
{{\mbox{div}}}(\lambda)=\mbox{trace}\{{\bf H}(\lambda)\}. \label{div_lambda}
\end{equation}

\subsection{The divergence in terms of $\rho$}

To develop the divergence in terms of $\rho$, we rely on the Demmler-Reinsch algorithm used in Eubank (1988). The algorithm can also speed up the computation of ${{\mbox{div}}}(\lambda)$.

{\it Demmler-Reinsch algorithm.} {Let ${\bf B}$ be a $d\times d$ matrix satisfying ${\bf B}^{-1}({\bf B}^{-1})'={\bf N}'{\bf N}$, where $N$ is $n\times d$. Let ${\bf U}$ be orthogonal and ${\bf C}$ be diagonal such that ${\bf U}{\bf C}{\bf U}'={\bf B}\bm\Omega_N {\bf B}'$. Define ${\bf Z}={\bf N}({\bf B}'{\bf U})$ and $\widehat{\bm\gamma}(\lambda)={\bf U}'({\bf B}^{-1})'\widehat{\bm\beta}(\lambda)$, where $\widehat{\bm\beta}(\lambda)$ is the solution to (\ref{new_penalty}). Then,  ${\bf N}\widehat{\bm\beta}(\lambda)={\bf Z}\widehat{\bm\gamma}(\lambda)$, $({\bf I}+\lambda {\bf C})\widehat{\bm\gamma}(\lambda)={\bf Z}'{\bf y}$, and $
\mbox{trace}\{{\bf H}(\lambda)\}=\sum_{j=1}^r(1+\lambda c_j)^{-1}$,
where $r$ is the rank of ${\bf C}$, and $c_1\geq c_2\geq \cdots \geq c_r$ are the non-zero diagonal elements of ${\bf C}$.}

Up to the choice of $\lambda$ and $\rho$, the problem (\ref{new_penalty}) is equivalent to
\begin{eqnarray}
\mbox{argmin\ }||\by-{\bf N}\bm\beta||^2_2 {\mbox{\ \ s.t. \ }}\bm\beta'\bm\Omega_N\bm\beta \leq \rho. \label{new_constraint}
\end{eqnarray}
Because $\bm\beta'{\bm\Omega_N}\bm\beta=\bm\gamma'{\bf C}\bm\gamma$, the problem (\ref{new_constraint}) becomes
\begin{eqnarray}
\mbox{argmin\ }||{\by}-{\bf Z}\bm\gamma||^2_2 \mbox{\ \ s.t.\ }\bm\gamma'{\bf C}\bm\gamma\leq\rho. \label{transformed_constraint}
\end{eqnarray}

Let $\widehat{\bm\gamma}(\rho)=(\widehat\gamma_1, \cdots, \widehat\gamma_d)'$ be the solution to the problem (\ref{transformed_constraint}) with constraint $\rho$ and $\widehat{\bm\gamma}^0$ be the solution to the problem without constraint. By some tedious arguments in Appendix, we derive the following divergence formula in terms of $\rho$.

\vskip 5pt

{\bf Theorem} {\it Following the notation in the description of Demmler-Reinsch algorithm,
\begin{eqnarray}
\mbox{{\mbox{div}}}(\rho)=(d-r)+\sum_{j=1}^{r-1}\frac{1}{1+\phi_j}, \label{div_rho}
\end{eqnarray}
where $\tau=||\hat{\bm\gamma}^0-\hat{\bm\gamma}(\rho)||_2$ and for $j=1, \cdots, r-1$,
$$\phi_j=\frac{\tau}{\sqrt{\sum_{l=1}^r\widehat{\gamma}_l^2c_l^2}}
\frac{c_{j}^2c_{j+1}\widehat{\gamma}_j^2+c_{j}c_{j+1}^2\widehat{\gamma}_{j+1}^2}
{c^2_{j}\widehat{\gamma}_{j}^2+c^2_{j+1}\widehat{\gamma}_{j+1}^2}.$$
}

\section{Some extensions}

The formula in the theorem can be extended to many other settings; for example, penalized splines, ridge regression, and functional linear regression. For these settings, ${\mbox{div}}(\lambda)$ is equal to the trace of the corresponding hat matrix.

\subsection{Penalized splines}

Penalized splines approach was proposed by Eilers and Marx (1996) to estimate $f(x)$ in (\ref{model}). For fixed order $p$ and knots $\kappa_1<\cdots<\kappa_K$, penalized splines approach finds a function of form $f(x)=\beta_0+\sum_{j=1}^p\beta_jx^j+\sum_{k=1}^K\beta_{p+k}(x-\kappa_k)_+^p$ that minimizes
\begin{equation}
\sum_{i=1}^n(y_i-f(x_i))^2+\lambda\sum_{k=1}^K\beta_{p+k}^2 {\mbox{\ \ or\ \ }} \sum_{i=1}^n(y_i-f(x_i))^2 {\mbox{\ \ s.t.\ }}\sum_{k=1}^K\beta_{p+k}^2<\rho. \label{penalized splines}
\end{equation}
An advantage of penalized splines over smoothing splines is that $K$ is much smaller than $n$. Ruppert (2002) claimed that the choice of $K$ is not important as long as it is large enough.

To apply the theorem to derive ${\mbox{div}}(\rho)$ in penalized splines, let ${\bf N}$ be an $n\times(p+K)$ matrix with the $i$th row being $(1, x_i, \cdots, x_i^p, (x_i-\kappa_1)^p_+, \cdots, (x_i-\kappa_K)_+^p),$ and let ${{\bm\Omega}_N}=\mbox{diag}\{{\bm{0}_{p+1}}, {\bm{1}_{K}}\}$, a $(p+K)\times (p+K)$ matrix.

\subsection{Ridge regression}

Ridge regression was proposed by Hoerl and Kennard (1970). Given data $\{({\bf x}_i, y_i)\in {\mathbb{R}}^p\times \mathbb{R}, i=1, \cdots, n\}$, the ridge regression finds a $\bbeta=(\beta_1, \cdots, \beta_p)'$ that minimizes
\begin{eqnarray}
\sum_{i=1}^n(y_i-\beta_0-\bbeta'{\bf x}_i)^2+\lambda\sum_{j=1}^p\beta_j^2 {\mbox{\ \ or\ \ }}\sum_{i=1}^n(y_i-\beta_0-\bbeta'{\bf x}_i)^2 {\mbox{\ \ s.t.\ }}\sum_{j=1}^p\beta_j^2\leq \rho. \label{ridge}
\end{eqnarray}

To apply the theorem to derive ${\mbox{div}}(\rho)$ in ridge regression, let ${\bf N}$ be an $n\times (p+1)$ matrix with the $i$th row being $(1, x_1, \cdots, x_p)$, and let ${\rm\Omega_N}=\mbox{diag}\{0, {\bm{1}_{p}}\}$, a $(p+1)\times (p+1)$ matrix.

\subsection{Functional linear regression}

Given data $\{(x_i(\cdot), y_i)\in \mathcal{L}_2([0,1])\times \mathbb{R}, i=1, \cdots, n\}$, consider functional linear regression,
\begin{eqnarray*}
y_i=\alpha_0+\int_0^1x_i(t)\beta_0(t)dt+\epsilon_i, \label{functional linear model}
\end{eqnarray*}
where $\beta_0$ is assumed to be in a Sobolev space of order $2$, $\mathcal{W}_2^2([0,1])$.
To estimate $f_0[x]=\alpha_0+\int_0^1x(t)\beta_0(t)dt$,  find one functional which minimizes
\begin{eqnarray}
\sum_{i=1}^n(y_i-f[x_i])^2+\lambda\int_0^1[\beta^{(2)}(t)]^2dt {\mbox{\ \ or\ \ }} \sum_{i=1}^n(y_i-f[x_i])^2{\mbox{\ \ s.t.\ }}\int_0^1[\beta^{(2)}(t)]^2dt\leq \rho, \label{Yuan RKHS approach}
\end{eqnarray}
among $\{f:\mathcal{L}_2([0, 1])\rightarrow\mathbb{R}\mid f[x]=\alpha+\int_0^1x(t)\beta(t)dt: \alpha\in\mathbb{R}, \beta\in\mathcal{W}_2^2([0,1])\}$.

For the functional linear regression, Yuan and Cai (2010) developed a reproducing kernel Hilbert space (RKHS) approach. They showed that the solution to (\ref{Yuan RKHS approach}) can be written as
\begin{eqnarray*}
\beta(t)=d_1+d_2t+\sum_{i=1}^nc_i\int_0^1[x_i(t)-\bar{x}(s)]K(t,s)ds,
\end{eqnarray*}
where $\bar{x}(s)=\sum x_i(s)/n$ and $K(t,s)$ is a kernel function. Let $\bm\Sigma$ be an $n\times n$ matrix where $\{\bm\Sigma\}_{ij}=\int\int[x_i(t)-\bar{x}(s)]K(t,s)[x_j(t)-\bar{x}(s)]dsdt$, ${\bf T}$ an $n\times 2$ matrix where $\{{\bf T}\}_{ij}=\int[x_i(t)-\bar{x}(t)]t^{j-1}dt$, ${\bf d}=(d_1, d_2)'$, and ${\bf c}=(c_1, \cdots, c_n)'$. Then problem (\ref{Yuan RKHS approach}) becomes
\begin{eqnarray*}
||{\bf y}-({\bf T}{\bf d}+{\bm\Sigma{\bf c}})||_2^2+\lambda{\bf c}'{\bm\Sigma}{\bf c} {\mbox{\ \ or \ \ }}
 ||{\bf y}-({\bf T}{\bf d}+{\bm\Sigma{\bf c}})||_2^2 \mbox{\ \ s.t. \ } {\bf c}'{\bm\Sigma}{\bf c}\leq \rho. \label{functional linear model constraint}
\end{eqnarray*}

To apply the theorem to derive ${\mbox{div}}(\rho)$ in functional linear regression, let the QR decomposition of ${\bf T}$ be $({\bf Q}_1 : {\bf Q}_2)({\bf R}' : {\bm{0}})'$ where ${\bf Q}_1$ is $n\times 2$, ${\bf Q}_2$ is $n\times (n-2)$, ${\bf Q}=({\bf Q}_1 : {\bf Q}_2)$ is orthogonal and ${\bf R}$ is upper triangular, with ${\bf T}'{\bf Q}_2=\bm{0}$. Since ${\bf T}'{\bf c}=0$, ${\bf c}$ must be in the column space of ${\bf Q}_2$, giving ${\bf c}={\bf Q}_2\bm\eta$ for some $\bm\eta$ an $n-2$ vector. Therefore, to apply the theorem, let ${\bf N}$ be replaced by $n\times n$ matrix $(\bm\Sigma{\bf Q}_2 : {\bf T})$ and let $\bm\Omega_N$ be replaced by $n\times n$ matrix $\mbox{diag}(\bm{0}, \bm\Sigma)$.

\section{Simulation studies}

In this section, we conduct a simulation study to verify the divergence formula in terms of $\rho$ for smoothing splines. We adopt the simulation setting in Gu (1998). On $x_i=(i-0.5)/100$, $i=1, \cdots, 100$, we generated 100 replicates of data from (\ref{model}) with $f(x)=1+3\sin(2\pi x-\pi)$ and $\sigma^2=1$. As in Gu (1998), for $\lambda$ on a fine grid of $\log_{10}n\lambda=(-5)(0.05)(-1)$, we calculated the solution to (\ref{penalty}), and then determined retrospectively the corresponding $\rho=\int_{0}^{1}\widehat{f}^{(2)}(x)dx$. This implies that the connection between $\lambda$ and $\rho$ is replicate-specific.

First, we compare the convergence formulae in terms of $\lambda$ and $\rho$ respectively. For each replicate, at each $\lambda$ in the grid, $\widehat{f}(x)$ was calculated, the corresponding $\rho$ is calculated, and then $\mbox{div}(\lambda)$ and $\mbox{div}(\rho)$ are calculated through (\ref{div_lambda}) and (\ref{div_rho}) respectively. For the first 10 replicates, the divergences are summarized in Figure 1. In the left panel, the curves of $\mbox{div}(\lambda)$ against $\lambda$ (they are identical) are drawn in red and the curves of $\mbox{div}(\rho)$ against $\lambda$ are drawn in blue. In the right panel, the curves of $\mbox{div}(\lambda)$ against $\rho$ are drawn in red and the curves of $\mbox{div}(\rho)$ against $\rho$ are drawn in blue. 

\begin{figure}[h]
\begin{center}\caption{Divergence Formula (blue for $\rho$ and red for $\lambda$)}
\includegraphics[width=6 in, height=3 in]{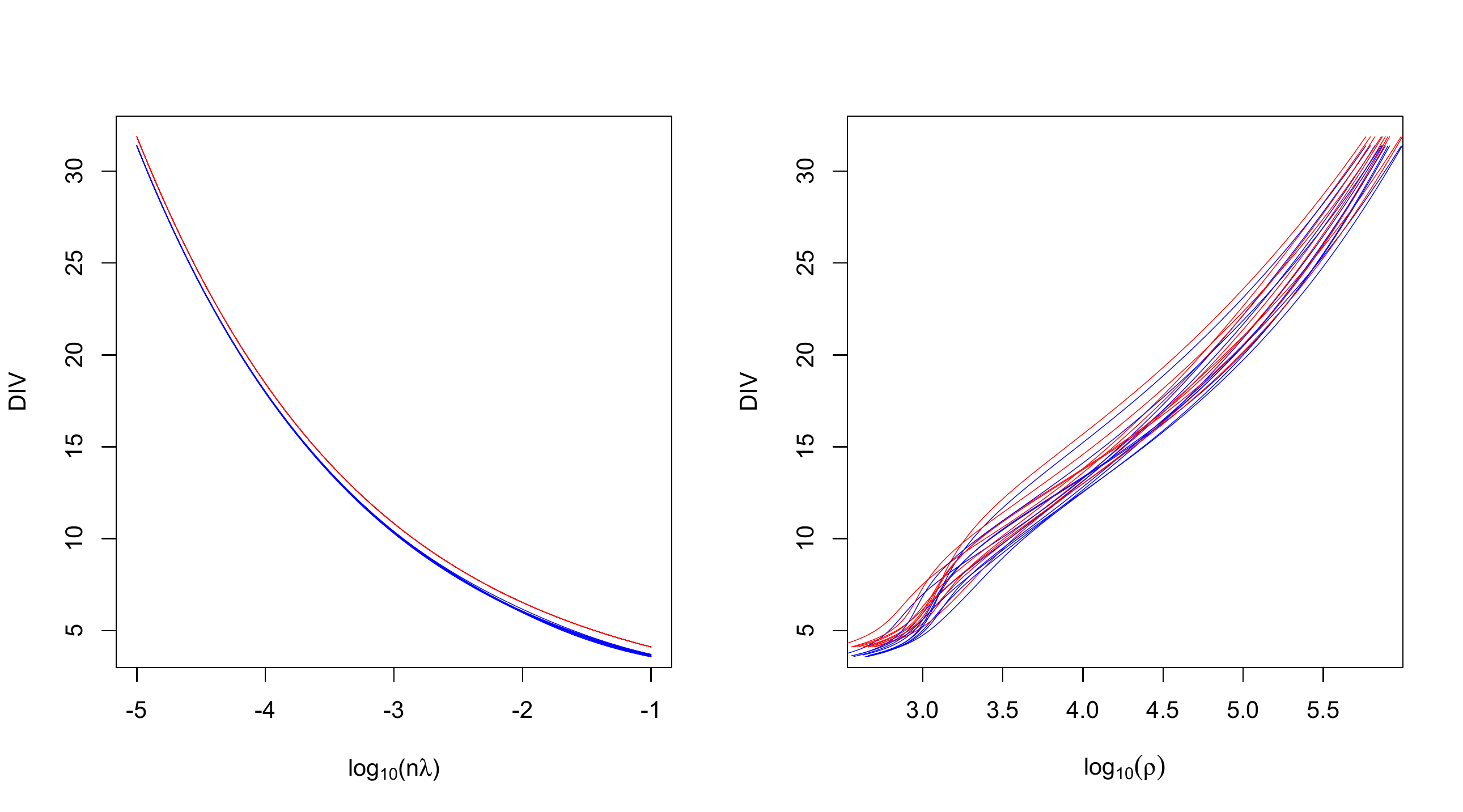}
\end{center}
\end{figure}

Further, as an application, the divergence formulas of $\lambda$ and $\rho$ can be used in construction of smoothing parameter selection criteria such as GCV and AIC. Again, because the criteria are commonly defined for both $\lambda$ and $\rho$, we use notation $\theta$, which could be either $\lambda$ or $\rho$. Let $\mbox{RSS}(\theta)=\sum_{i=1}^n[y_i-\widehat{\mu}_i(\theta)]^2$ be the residual sum of squres. An effective smoothing parameter selection criterion is Akaike Information Criterion (AIC; Akaike (1973)),
\begin{eqnarray}
\mbox{AIC}(\theta)=\log\mbox{RSS}(\theta)+2\mbox{div}(\theta).
\end{eqnarray}
Another effective smoothing parameter selection criterion is Generalized Cross-Validation (GCV; Craven and Wahba (1979)),
\begin{eqnarray}
\mbox{GCV}(\theta)=\frac{\mbox{RSS}(\theta)}{(n-\mbox{div}(\theta))^2}.
\end{eqnarray}

We compare the performances of AIC and GCV in terms of $\lambda$ and $\rho$. Following Caution 1 in Gu (1998), we consider the risk function indexed in terms of $\rho$,
$\mbox{Risk}(\rho)=E\frac{1}{n}\sum_{i=1}^n(\widehat{f}(x_i)-f(x_i))^2$,
where the expectation is with respect to $\epsilon_i$. By Caution 1, the risk function indexed in terms of $\lambda$ is meaningless because model index $\lambda$ is data-specific. The comparison is based on the following relative error in Hastie {\it et al.}~(2009, p.241),
\begin{eqnarray}
100\times\frac{\mbox{Risk}(\widehat{\rho})-\min_{\rho}\mbox{Risk}(\rho)}{\max_{\rho}\mbox{Risk}(\rho)-\min_{\rho}\mbox{Risk}(\rho)}.
\end{eqnarray}
We should explain $\widehat{\rho}$ in the above formula. If $\mbox{AIC}(\rho)$ is applied, $\widehat{\rho}=\arg\min_{\rho} \mbox{AIC}(\rho)$. If $\mbox{AIC}(\lambda)$ is applied, $\widehat{\rho}$ is defined as the counterpart of $\widehat{\lambda}=\arg\min_{\lambda}\mbox{AIC}(\lambda)$. Similar $\widehat{\rho}$ is defined for GCV. The results are summarized in Figure 2. It is found that the performances of criteria in terms of $\rho$ are almost the same as those in terms of $\lambda$. This finding supports the correctness of the derived formula. 

\begin{figure}[h]
\begin{center}\caption{Relative Error}
\includegraphics[width=6 in, height=3 in]{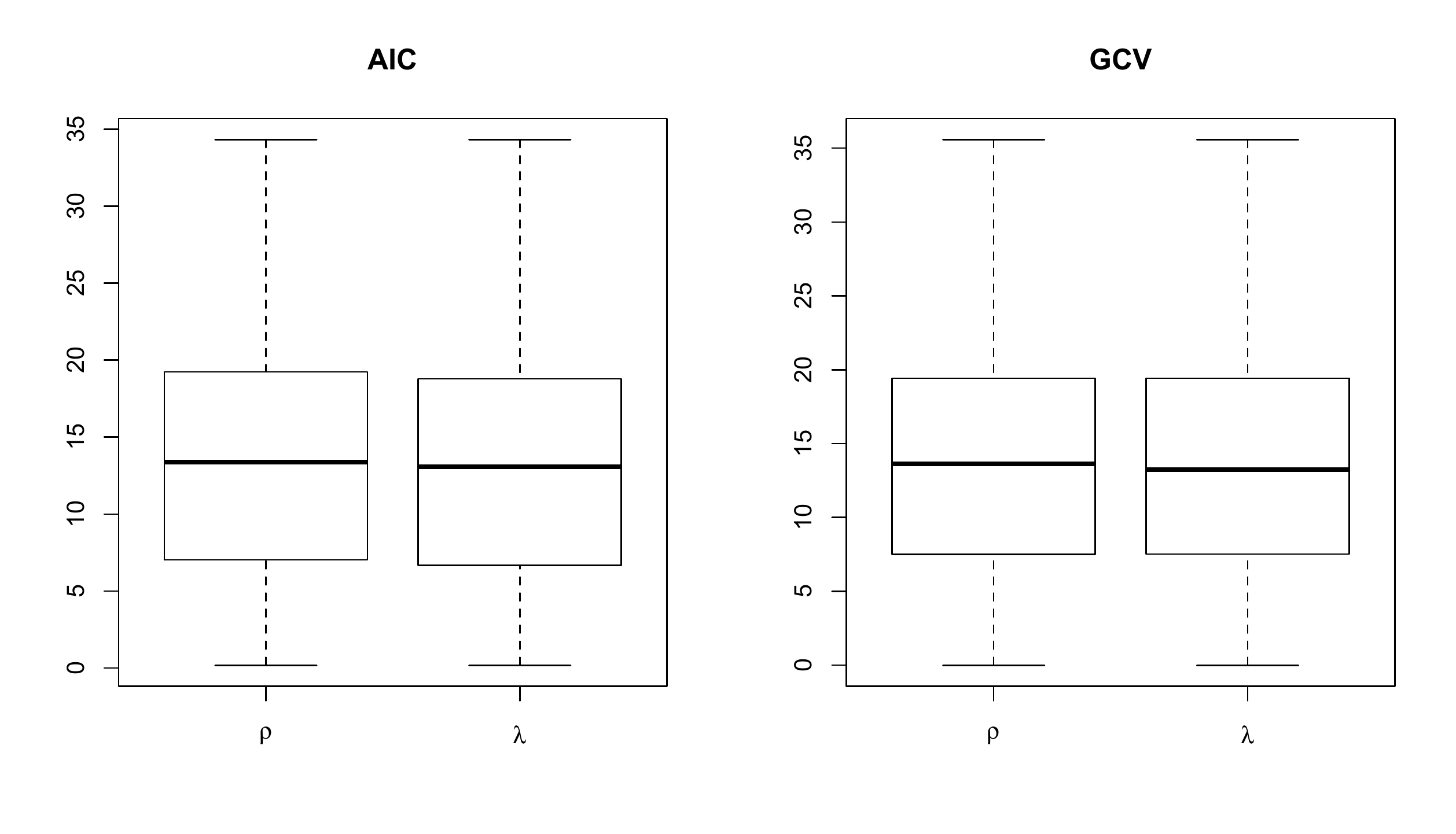}
\end{center}
\end{figure}

\section{Discussion}

For regularization methods with an $l_1$ penalty, the divergence formula (in terms of $\lambda$, if adopt our notation) was derived in Zou {\it et al.}~(2007). For regularization methods with an $l_1$ constraint, the divergence formula (in terms of $\rho$, if adopt our notation) was derived in Kato (2009). This manuscript considers the divergence formula for the $l_2$ regularization, which appears ahead of the $l_1$ regularization. 

Although the divergence formula (in terms of $\lambda$) for regularization methods with an $l_2$ penalty has been existing long ago, the divergence formula (in terms of $\rho$) for regularization methods with an $l_2$ is still not available in the literature. Now this missing formula is derived.

%
%

\begin{center}
{\bf\huge {Appendix}}
\end{center}
\renewcommand{\theequation}{A.\arabic{equation}}
\setcounter{equation}{0}

In this appendix, following Kato (2009), we develop the divergence formula in terms of $\rho$. On the boundary $\Omega=\{\bgamma\in R^d: \sum_{j=1}^rc_j\gamma_j^2=\rho\}$, $\bgamma$ can be transformed into polar coordinates: $\bgamma=(\sqrt{\rho}u(\theta_1, \cdots, \theta_{r-1}), \gamma_{r+1}, \cdots, \gamma_{d})'$, with $u(\theta_j, \cdots, \theta_{r-1})$ defined as
$$(\frac{\cos\theta_j}{\sqrt{c_j}}, \frac{\sin\theta_j\cos\theta_{j+1}}{\sqrt{c_{j+1}}}, \cdots, \frac{\sin\theta_j\sin\theta_{j+1}\cdots\cos\theta_{r-1}}{\sqrt{c_{r-1}}}, \frac{\sin\theta_j\sin\theta_{j+1}\cdots\sin\theta_{r-1}}{\sqrt{c_r}})',$$
where $0\leq\theta_j\leq\pi$, $j=1, \cdots, r-2$, and $0\leq\theta_{r-1}\leq 2\pi$. On the boundary, which is in a $(d-1)$-dim smooth minifold, the partial derivative of $\bgamma$ with respect to $\theta_j$ is given by
$$\frac{\partial\bgamma}{\partial\theta_j}=\sqrt{\rho}\sin\theta_1\cdots\sin\theta_{j-1}({\bf 0}'_{j-1}, v(\theta_j, \cdots, \theta_{r-1}), {\bf 0}'_{d-r} )',$$
with ${\bf 0}_{l}$ defined as $l$-dim vector of all components being zero, and $v(\theta_j, \cdots, \theta_{r-1})$ as  $$(-\frac{\sin\theta_j}{\sqrt{c_j}}, \frac{\cos\theta_j\cos\theta_{j+1}}{\sqrt{c_{j+1}}}, \cdots, \frac{\cos\theta_j\sin\theta_{j+1}\cdots\cos\theta_{r-1}}{\sqrt{c_{r-1}}}, \frac{\cos\theta_j\sin\theta_{j+1}\cdots\sin\theta_{r-1}}{\sqrt{c_r}})'.$$
Furthermore, on the boundary, the second partial derivatives, for $1\leq j<k\leq r$, are
\begin{eqnarray*}
\frac{\partial^2\bgamma}{\partial\theta_j^2}&=&-\sqrt{\rho}\sin\theta_1\cdots\sin\theta_{j-1}({\bf 0}'_{j-1}, u(\theta_j, \cdots, \theta_{r-1}), {\bf 0}'_{d-r})',\\
\frac{\partial^2\bgamma}{\partial\theta_j\partial\theta_k}&=&
\sqrt{\rho}\sin\theta_1\cdots\sin\theta_{j-1}\cos\theta_j\sin\theta_{j+1}\cdots\sin\theta_{k-1}
({\bf 0}'_{k-1}, u(\theta_k, \cdots, \theta_{r-1}), {\bf 0}'_{d-r})'.
\end{eqnarray*}
Define as $\bnu$ the following vector which is orthogonal to the tangent space of $\Omega$ at $\bgamma$,
$$(\sqrt{c_1}\cos\theta_1, \sqrt{c_2}\sin\theta_1\cos\theta_2, \cdots, \sqrt{c_{r-1}}\sin\theta_1\sin\theta_2\cdots\cos\theta_{r-1}, \sqrt{c_r}\sin\theta_1\cdots\sin\theta_{r-1}, {\bf 0}'_{d-r})',$$
and $\bnu_0=\bnu/||\bnu||_2$.

We are ready to calculate the {\it first fundamental form} and {\it second fundament form} defined in Kato (2009, p.1342-1343). For this aim, let $\btheta=(\theta_1, \cdots, \theta_{r-1})'$ and $\bomega=(\gamma_{r+1}, \cdots, \gamma_d)'$. The first fundamental form equals, noting that $\frac{\partial\bgamma'}{\partial\bomega}\frac{\partial\bgamma}{\partial\bomega'}=I_{d-r}$ and $\frac{\partial\bgamma'}{\partial\btheta}\frac{\partial\bgamma}{\partial\bomega'}={\bf 0}_{(r-1)\times(d-r)}$,
$$G=\mbox{diag}(G_{11}, I_{d-r}),$$
where $G_{11}=\frac{\partial\bgamma'}{\partial\btheta}\frac{\partial\bgamma}{\partial\btheta'}=L'\mbox{diag}^{-1}(c_1, \cdots, c_r)L$, with $L$ being the followng $r\times(r-1)$ matrix,
$$\sqrt{\rho}[v^0(\theta_1, \cdots, \theta_{r-1}), \cdots,
\prod_{l=1}^{j-1}\sin\theta_l\left(
                                  \begin{array}{c}
                                  {\bf 0}_{j-1}\\
                                  v^0(\theta_j,\cdots,\theta_{r-1}) \\
                                  \end{array}
                                  \right), \cdots,
\prod_{l=1}^{r-2}\sin\theta_l\left(
                                  \begin{array}{c}
                                  {\bf 0}_{r-2}\\
                                  v^0(\theta_{r-1}) \\
                                  \end{array}
                                  \right)],$$
and $v^0(\theta_j, \cdots, \theta_{r-1})$ being the following $r-j+1$ vector,
$$(-{\sin\theta_j}, {\cos\theta_j\cos\theta_{j+1}}, \cdots, {\cos\theta_j\prod_{l=j+1}^{r-2}\sin\theta_l\cos\theta_{r-1}}, {\cos\theta_j\prod_{l=j+1}^{r-1}\sin\theta_l})'.$$

\noindent The second fundamental form equals, noting that $\frac{\partial^2\bgamma}{\partial\btheta\partial\bomega'}={\bf 0}_{(r-1)\times(d-r)}$ and $\frac{\partial^2\bgamma}{\partial\omega\partial\bomega'}={\bf 0}_{(d-r)\times(d-r)}$,
$$H=\mbox{diag}(H_{11}, {\bf 0}_{(d-r)\times (d-r)}),$$
where $H_{11}$ is a $(r-1)\times (r-1)$ matrix with the $(j, k)$ component being $-\tau <\bnu, \frac{\partial^2\bgamma}{\partial\theta_j\partial\theta_k}>$. Here $<\cdot,\cdot>$ is the ordinary Euclidean inner product in $R^d$. It can be verified that
$H_{11}=\tau\mbox{diag}(h_1, \cdots, h_{r-1})=\frac{\tau}{\sqrt{\rho}||\bnu||_2}L'L$, where $h_j=\sqrt{\rho}\sin^2\theta_1\cdots\sin^2\theta_{j-1}/||\bnu||_2$.

By Lemma 3.2 in Kato (2009), we can obtain the divergence formula in terms of $\rho$,
$$\mbox{div}(\rho)=(d-r)+\sum_{j=1}^{r-1}\frac{1}{1+\phi_j},$$
where $\phi_j$, $j=1, \cdots, r-1$, are the eigenvalues satisfying the equation
$$\mbox{det}(H_{11}-\phi G_{11})=0.$$
To find the eigenvalues, note that $H_{11}-\phi G_{11}=L'\mbox{diag}(d_1,\cdots,d_r)L$, where $d_j=\frac{\tau}{\sqrt{\rho}||\bnu||_2}-\frac{\phi}{c_j}$.
It can be verified the $j$th diagonal component of $L'\mbox{diag}(d_1,\cdots,d_r)L$ equals
$$e_j(\phi)=\prod_{l=1}^{j-2}\sin^2\theta_l(d_{j-1}\sin\theta_{j-1}+d_j\cos\theta_{j-1}/\cos^2\theta_j),$$
for $j=1, \cdots, r-1$. Let $\phi_j$ be the solution to the equation $e_j(\phi)=0$, $j=1, \cdots, r-1$. Since matrix $L'\mbox{diag}(d_1,\cdots,d_r)L$ is non-negative definite, we can conclude that $\phi_j$, $j=1, \cdots, r-1$, are the eigenvalues we need. It is easy to see that $\phi_j$ is the solution to $e_j(\phi)=0$, and therefore the divergence formula in terms of $\rho$ is obtained. $\Box$

\end{document}